\documentclass[
    a4paper,
    10pt,
    oneside,
]{article}

\usepackage{float}
\usepackage[table]{xcolor}
\usepackage{orcidlink}
\usepackage{url}
\usepackage[utf8]{inputenc}
\usepackage{amsmath,amssymb,amsfonts}
\usepackage{mathtools}
\usepackage{graphicx}
\usepackage{textcomp}
\usepackage[T1]{fontenc} \usepackage{enumitem} \usepackage[most,many,breakable]{tcolorbox}
\usepackage{multicol}
\usepackage{multirow}
\usepackage{wrapfig}
\usepackage{todonotes}
\usepackage{subcaption}

\usepackage[
    backend=biber,		bibwarn=true,
	bibencoding=utf8,	sorting=none,
	bibstyle=ieee,
	citestyle=ieee-comp,
isbn=false,
	url=false,
	doi=false,
	eprint=false,
	maxnames=3,
]{biblatex}

\definecolor{SIEMENS_PETROL}{HTML}{009999}
\definecolor{LIGHT_PETROL}{HTML}{00c1b6}
\definecolor{BOLD_GREEN}{HTML}{00ffb9}
\definecolor{SOFT_GREEN}{HTML}{00d7a0}
\definecolor{BOLD_BLUE}{HTML}{00e6dc}
\definecolor{SOFT_BLUE}{HTML}{00bedc}
\definecolor{DEEP_BLUE}{HTML}{000028}
\definecolor{LIGHT_SAND}{HTML}{f3f3f0}
\definecolor{DARK_SAND}{HTML}{aaaa96}
\definecolor{SOFT_SAND}{HTML}{c5c5b8}
\definecolor{BRIGHT_SAND}{HTML}{dfdfd9}
\definecolor{DARK_YELLOW}{HTML}{f7c600}
\definecolor{YELLOW}{HTML}{ffd732}
\definecolor{SOFT_YELLOW}{HTML}{ffe270}
\definecolor{DARK_GREEN}{HTML}{00646e}
\definecolor{GREEN}{HTML}{00af8e}
\definecolor{DARK_PURPLE}{HTML}{553ba3}
\definecolor{PURPLE}{HTML}{805cff}
\definecolor{SOFT_PURPLE}{HTML}{b4a8ff}
\definecolor{RED}{HTML}{ef0137}
\definecolor{DARK_ORANGE}{HTML}{ec6602}
\definecolor{ORANGE}{HTML}{ff9000}

\newcommand*\circled[1]{\tikz[baseline=(char.base)]{
		\node[shape=circle,draw,inner sep=1pt] (char) {#1};}}

\newcommand{\circlenum}[1]{{\tiny\small\protect\circled{#1}}}
 
\usepackage[
  left=3.5cm, 
  right=3.5cm,
  top=3cm,
  bottom=3.5cm,
]{geometry}

\makeatletter
\let\orig@xfloat\@xfloat
\def\@xfloat#1[#2]{\orig@xfloat{#1}[htbp]}
\makeatother

\newenvironment{IEEEkeywords}{\begin{quote}\small \textbf{Keywords: }}{\end{quote}}
\renewenvironment{abstract}{\begin{quote}\small \textbf{Abstract: }}{\end{quote}}

\hypersetup{pdftitle={Efficient Quantile-Resolved Hosting Capacity Assessment on Nodal Level for Low-Voltage Grids},
	pdfauthor={Maximilian Köhler, Edwin Mora, Mathias Duckheim, Stefan Niessen},
	pdfsubject={Preprint from a Conference Paper to IEEE PES ISGT Europe 2026},
	pdfcreator={Copyright (C) 2026 IEEE.  All rights reserved.},
	pdfpagemode=UseOutlines, 		pdfdisplaydoctitle=true, 		}

\begin{document}

\title{
    Efficient Quantile-Resolved Hosting Capacity Assessment on Nodal Level for Low-Voltage Grids
    \thanks{This work was conducted within the scope of the research project GridAssist and was supported through the “OptiNetD” funding initiative by the German Federal Ministry for Economic Affairs and Energy (BMWE) as part of the 8th Energy Research Programme.}
    \thanks{
      \copyright~2026 IEEE.
      Personal use of this material is permitted.
      Permission from IEEE must be obtained for all other uses, in any current or future media, including reprinting/republishing this material for advertising or promotional purposes, creating new collective works, for resale or redistribution to servers or lists, or reuse of any copyrighted component of this work in other works.
      This work was submitted and accepted at the conference IEEE PES Innovative Smart Grid Technologies 2026 in Budapest, Hungary.
      The original publication will be available at IEEE Xplore, this document is a pre-print version. }
}
\author{
  Maximilian Köhler\(^{*1,2}\), Edwin Mora\(^2\),\\Mathias Duckheim\(^2\), Stefan Niessen\(^{1,2}\)\\
  \small \textsuperscript{1}\textit{Technical University of Darmstadt}, Darmstadt, Germany \\
  \small \textsuperscript{2}\textit{Siemens AG}, Erlangen, Germany \\
  \small \(*\)koehler.maximilian@siemens.com
}
\date{
}

\maketitle

\begin{abstract}
    Hosting capacity—the maximum additional capacity a network can accommodate without violating operational limits—is a key metric in distribution system planning and operation.
    Decisions on grid reinforcements and the deployment of flexibility management require not only the worst-case HC but also an understanding of the distribution of HC under different likelihoods in load and generation patterns.
    Quantile-resolved HC distributions provide this view by expressing HC as a function of an acceptable operational limit exceedance likelihood.
    Monte Carlo sampling is the established approach for computing such distributions but demands large computational resources.
    Approximations sacrifice either accuracy, the ability to capture uncertainty correlations, or scalability when assessing real-world networks.
    This paper introduces a computationally efficient method for calculating distributions for quantile-resolved HC.
    It uses a representation of load samples as multivariate normal distribution, propagated through a linearized power flow model.
    This allows for leveraging a re-parametrized AC-OPF problem for each hosting capacity quantile.
    Benchmarking against Monte Carlo-based methods on realistic LV networks demonstrates that the proposed method achieves comparable accuracy with a mean deviation of approx. 3\,\%, while reducing computational time by orders of magnitude.
    For the exemplary networks the computational time decreases from 11\,min to 2\,s, and 38\,h to 50\,s, respectively.
    The method's scalability is also suitable for recalculation in 15-minute cycles encountered in DSO practice for e.g., real-time grid management.
\end{abstract}


\section{Introduction}
\label{sec:introduction}

Distribution System Operators (DSOs) face increasing challenges managing Distributed Energy Resources (DERs) at the Low-Voltage (LV) level \cite{vnbdigital_2023}.
Traditional worst-case planning approaches, which assume simultaneous peak loading across all DERs, ensure operational safety but lead to systematic over-investment.
Scenario-based planning that accounts for load and generation diversity can reduce grid reinforcement costs by up to 45\% \cite{andreae_2023}, necessitating new methods for uncertainty-aware capacity assessment.

Hosting capacity (HC) quantifies the maximum capacity that can be connected at a network node without violating operational constraints, serving as a key metric for both planning and operational decisions.
For uncertainty-aware planning, quantile-resolved HC—which provides capacity limits at different risk levels e.g., 95th, 99th percentile—enables risk-based decision-making rather than conservative worst-case assessments.
Computing quantile-resolved HC requires evaluating numerous load and generation scenarios.
Sample-based methods, such as Monte Carlo sampling, solve an Optimal Power Flow (OPF) for each scenario to construct the HC distribution, providing accurate results at the cost of high computational effort.
This computational challenge is prohibitive for operational applications requiring periodic calculations at e.g., 15-minute cycles, or in large-scale grid planning.

Existing approaches propose trade-offs between accuracy and scalability.
Approximation methods such as polynomial chaos expansion (PCE), surrogate models, and machine learning \cite{islam_2023, islam_2025, geng_2021, madavan_2024, koirala_2020} reduce computational effort but introduce significant approximation errors or fail to preserve load-generation correlations. 
Robust optimization methods \cite{muhlpfordt_2018} provide conservative bounds but do not yield quantile-resolved distributions needed for risk-aware planning.
This paper introduces probabilistic hosting capacity assessment (pHCA), which computes quantile-resolved HC distributions through a single AC-OPF formulation that explicitly incorporates load and generation uncertainties and their correlations.
Unlike iterative sampling methods, pHCA determines the complete HC distribution in one optimization, achieving comparable accuracy to Monte Carlo sampling.

The remainder of the paper is organized as follows: Section~\ref{sec:modeling} models the problem mathematically and introduces metrics to compare the resulting distributions.
Section~\ref{sec:results} compares the results of sHCA and pHCA with respect to result quality and computational efficiency.
An application of the pHCA on a realistic LV grid demonstrates the computational efficiency and accuracy of the method.
Section~\ref{sec:discussion} discusses these results.
Finally, Section~\ref{sec:summary} summarizes and gives an outlook on future work and applications.

\section{Problem Formulation and Solution Approaches}
\label{sec:modeling}

This section introduces the mathematical formulations for the grid representation and the hosting capacity assessment problem.
It is divided into grid modeling, formulation of the benchmark calculation algorithm, formulation of our pHCA, and the definition of evaluation metrics.

\subsection{Grid Modeling}
We represent the power network as a set of nodes \(\mathcal{N}\) and edges \(\mathcal{E}\), which represent busbars and branch elements, respectively.
\(\mathcal{N}'\) is the set of non-slack nodes.
In addition, let \(\mathcal{D}\) be the set of loads and generators connected to the network.
The matrix of active and reactive powers associated to the load and generator elements is defined as
\begin{align*}
    \mathbf{W}_\mathrm{load} =& \begin{bmatrix}
        p_{0, 0} & \cdots & p_{0, t} & \cdots & p_{0, \vert \mathcal{T} \vert} \\
        q_{0, 0} & \cdots & q_{0, t} & \cdots & q_{0, \vert \mathcal{T} \vert} \\
        \vdots &  & \ddots &  & \vdots \\
        p_{\vert \mathcal{D} \vert, 0} & \cdots & p_{\vert \mathcal{D} \vert, t} & \cdots & p_{\vert \mathcal{D} \vert, \vert \mathcal{T} \vert} \\
        q_{\vert \mathcal{D} \vert, 0} & \cdots & q_{\vert \mathcal{D} \vert, t} & \cdots & q_{\vert \mathcal{D} \vert, \vert \mathcal{T} \vert}
    \end{bmatrix}, \in \mathbb{R}^{2 \vert \mathcal{D} \vert \times \vert \mathcal{T} \vert}
\end{align*}
It considers samples \(t\) in a selectable sample set \(\mathcal{T}\) and is referred to as the reference load hypothesis.
We denote the \(t\)-th load sample of \(\mathbf{W}_\mathrm{load}\) as \(\mathbf{w}_t\).
Generators are modeled as negative loads.
The nodal loads \(\mathbf{W}_\mathrm{node} \in \mathbb{R}^{2 \vert \mathcal{N}' \vert \times \vert \mathcal{T} \vert}\) are obtained by 
\begin{align*} 
    \mathbf{W}_\mathrm{node} &= \mathbf{C} \mathbf{W}_\mathrm{load},
\end{align*}
with a load-node incidence matrix \(\mathbf{C} \in \mathbb{R}^{\vert \mathcal{N}' \vert \times \vert \mathcal{D} \vert}\).
Multiple DERs may be connected to the same node, but not every node needs to have a connected load or generator.
We model the power flow equations as a function \(y = h(\mathbf{w}_{\mathrm{node},t})\) of the power at each node and sample.
Thus, for a given \(\mathbf{w}_{\mathrm{node}}\), we can derive the electrical state of the network, i.e.,
\begin{align*}
	\mathbf{y} = \begin{bmatrix} \mathbf{v} \\ \mathbf{i}  \end{bmatrix}, \quad \text{with } \mathbf{v} \in \mathbb{R}^{\vert \mathcal{N}' \vert},
    ~\mathbf{i} \in \mathbb{R}^{\vert \mathcal{E} \vert},
\end{align*}
where \(\mathbf{v}\) is the vector of voltage magnitudes at all (non-slack) nodes and \(\mathbf{i}\) is the vector of all current magnitudes. 
All samples of \(\mathbf{W}_\mathrm{load}\) can be obtained from historical data or load forecasts, or sampled from a normal distribution with mean \(\mu\) and covariance \(\sigma^2\) as follows:
\begin{align*} 
    \mathbf{W}_\mathrm{dist} \sim \boldsymbol{\mathcal{N}}(\mu, \sigma^2)
\end{align*}

\subsection{Hosting Capacity Analysis as Optimization Problem}
HCA can be formulated as a mathematical optimization task which, for node \(n\) and sample \(t\), maximizes the load  \(u_{n,t}\) that can be added to that node, such that the voltage and currents are feasible.
The safe operation of the network with respect to operational limits for voltages and currents is modeled using constraints of the form \(g(\mathbf{y}) \leq 0\), with
\begin{align}
    g(\mathbf{y}) &= \mathbf{A} \mathbf{y}  - \mathbf{b}, \label{eq:constraints}
\end{align}
and
\begin{align*}
    \mathbf{A} &= \begin{bmatrix}
        - \mathbf{I} & 0 \\
        + \mathbf{I} & 0 \\
        0 & +\mathbf{I}
    \end{bmatrix}, \\[10pt]
    \mathbf{b} &= \begin{bmatrix}
        \mathbf{v}_\mathrm{min} \\
        - \mathbf{v}_\mathrm{max} \\
        - \mathbf{i}_\mathrm{max}
    \end{bmatrix},~\mathbf{v}_\mathrm{min}, \mathbf{v}_\mathrm{max} \in \mathbb{R}^{\vert \mathcal{N}' \vert},~\mathbf{i}_\mathrm{max} \in \mathbb{R}^{\vert \mathcal{E} \vert},
\end{align*}
where \(\mathbf{I}\) represents an identity matrix of appropriate dimensions.
In this work we use the voltage limits \(v_{\mathrm{min},n} = 0.9\) p.u. and \(v_{\mathrm{max},n} = 1.1\) p.u., and the current limit \(i_{\mathrm{max},n}\) is set to the maximum current of the respective branch according to the network data.

\subsubsection{Sample-based Hosting Capacity Assessment (sHCA)}\label{subsec:det-hca}
The quantile-resolved HCA is calculated by solving the following optimization problem \(\forall~n \in \vert \mathcal{N}' \vert\), \(\forall~t \in \vert \mathcal{T} \vert\):
\begin{align}
    u^*_{n, t} &= \underset{u_n \in \mathbb{R}}{\operatorname{argmax}} \quad  u_n \label{eq:bopf-optimization}\\
    & \text{s.t.} \quad g(h(\mathbf{C} \mathbf{w}_{\mathrm{load},t} + \mathbf{e}_n u_n)) \leq 0, \notag 
\end{align}
where \(u_n\) is the additional load at node \(n\), \(\mathbf{w}_{\mathrm{load},t}\) is the load hypothesis at the sample \(t\) excluding the added DER capacity and \(\mathbf{e}_n\) is a unit vector selecting the node at which the HC load is added.
The set 
\begin{align*}
    \mathcal{U}_n = \{u^*_{n,t}~\vert~\forall~t \in \mathcal{T}\}
\end{align*}
is interpreted as the distribution of the HC at node \(n\) for the given load hypothesis.
It represents all samples of the HC distribution at node \(n\).
From the distribution function of the set, a quantile \(x \in [0,1]\) of the hosting capacity can be determined with the quantile function \(F_{\mathcal{U}_n}^{-1}\) or inverse cumulative distribution function (CDF), i.e.,
\begin{align*}
    \text{HC}_{x,n} = F_{\mathcal{U}_n}^{-1}(x), ~\forall~n \in \mathcal{N}'.
\end{align*}

As an interpretation example one could have a \(\text{HC}_{0.05,n}= 10 \,\mathrm{kW}\) for a quantile of 5\,\% at node \(n\).
This means that in 95\,\% of the cases, an additional 10 kW can be connected at node \(n\) without exceeding any operational limits of the grid.
The sHCA approach requires solving \(\vert \mathcal{T} \vert\) optimization problems for each node \(n\), leading to \(\vert \mathcal{T} \vert \times \vert \mathcal{N}' \vert\) OPF calculations for the quantile-resolved HCA.
This is computationally expensive, especially for large networks and large number of samples, which motivates the development of the more efficient algorithm proposed next.

\subsubsection{Probabilistic Hosting Capacity Assessment (pHCA)}\label{subsec:prob-hca}

The quantile-resolved HCA with the pHCA consists of the steps:
\begin{enumerate}[label=\bfseries\tiny\small\protect\circled{\arabic*}]
    \item Generate a distribution \(\mathbf{W}_\mathrm{dist}\) for the load hypothesis,
    \item Determine distributions \(\mathbf{y}_\mathrm{dist}\) for the electrical state of the network,
    \item Calculate marginalized distributions for voltages and currents \(\mathbf{v}_\mathrm{dist}, \mathbf{i}_\mathrm{dist}\),
    \item Determine tightened constraints similar to Eq.~(\ref{eq:constraints}),
    \item Calculate approximate for HC \(\hat{u}_{n,x}~\forall~n \in \mathcal{N}', x \in [0,1]\).
\end{enumerate}
For the steps \circlenum{1} to \circlenum{3}, earlier publications describe the calculation of voltage and current distributions based on nodal power distributions and a linearized power flow model \cite{arefi_2015,buchta_2023}.
In step \circlenum{1}, the load hypothesis is defined by expected values and covariances for the nodal powers, which can be based on historical timeseries data or load forecasts.
Step \circlenum{2} determines the state distributions, voltage and current amplitudes, via a linearized power flow model on the distributions of the load hypothesis.
In step \circlenum{3}, marginalized distributions for the voltages \(\mathbf{v}_\mathrm{dist}\) and currents \(\mathbf{i}_\mathrm{dist}\) are calculated.
These are leveraged in step \circlenum{4} for the reformulation of the constraints, where the voltage and current amplitude differences \(\Delta \mathbf{v}(x) \in \mathbb{R}^{\vert \mathcal{N}' \vert}\) and \(\Delta \mathbf{i}(x) \in \mathbb{R}^{\vert \mathcal{E} \vert}\) are calculated with the quantile function as follows:
\begin{align*}
    \Delta \mathbf{v}(x) &= \mu[\mathbf{v}_\mathrm{dist}] - F_{\mathbf{v}_\mathrm{dist}}^{-1}(x), \\
    \Delta \mathbf{i}(x) &= \mu[\mathbf{i}_\mathrm{dist}] - F_{\mathbf{i}_\mathrm{dist}}^{-1}(x),
\end{align*}
The new constraints \(g'(\cdot)\) are then defined in the same manner as \(g(\cdot)\) but with the voltage limits tightened in the vector \(\mathbf{b}'\) as follows:
\begin{align*}
    \mathbf{b}' = \begin{bmatrix}
        \mathbf{v}_\mathrm{min} + \Delta \mathbf{v}(x) \\
        - \mathbf{v}_\mathrm{max} + \Delta \mathbf{v}(x) \\
        - \mathbf{i}_\mathrm{max} + \Delta \mathbf{i}(x)
    \end{bmatrix}
\end{align*}
For the pHCA, the optimization problem becomes a reparametrized OPF problem, where the expected values of the additional nodal power are optimized, and the constraints are defined by the tightened grid limitations \(g'(\cdot)\) as follows:
\begin{alignat}{3}
	\hat{u}_{n,x} & = \underset{u}{\operatorname{argmax}} \quad u \label{eq:pgm-optimization}\\
	&\text{s.t.} \quad g'(h(\mathbf{C} \times \mu[\mathbf{w}_\mathrm{dist}] + \mathbf{e}_n u)) \leq 0 \notag
\end{alignat}
This optimization problem is solved with full AC power flow equations, the linearization is only used for the calculation of the state distributions and the reformulation of the constraints.
For the proposed definition of probabilistic Hosting Capacity, we approximate the equivalence of both methods as
\begin{align*}
    \hat{\text{HC}}_{n,x} &\approx \text{HC}_{x,n}, \quad \forall n \in \mathcal{N}', x \in [0,1].
\end{align*}
To calculate results for a given network, the number of OPF calculations is reduced to \(\vert \mathcal{N}' \vert\) for the probabilistic optimization approach.

\subsection{Evaluation Metrics}
\label{subsec:evaluation-metrics}

In order to compare the results of the two approaches, we use two metrics to quantify the similarity of the resulting HC distributions.
The Mean Relative Error (\(\text{MRE}_n\)) and the Quantile Relative Error (\(\text{qRE}_{n,x}\)) for node \(n\) and quantile \(x\) are defined as follows:
\begin{align*}
    \text{MRE}_n &= \frac{\mu[\hat{\text{HC}_n}] - \mu[\text{HC}_n]}{\mu[\text{HC}_n]}, \\
    \text{qRE}_{n,x} &= \frac{\vert \hat{\text{HC}}_{n,x} - \text{HC}_{n,x} \vert}{\sigma[\text{HC}_n]}, \\
    \text{with } x &= 0.05, \text{ for all nodes } n \in \mathcal{N}' \notag
\end{align*}
Here, $\mu[\hat{\text{HC}_n}]$ and $\mu[\text{HC}_n]$ denote the expected values of the HC distribution at node $n$ for the pHCA and sHCA approaches, respectively.
$\sigma[\text{HC}_n]$ represents the standard deviation of the HC distribution obtained from the sHCA approach.

The \(\text{MRE}_n\) quantifies the expectation bias of the pHCA distribution with respect to the sHCA distribution.
The \(\text{qRE}_{n,x}\) quantifies the discrepancy of the \(x\)-th quantile at node \(n\) obtained via pHCA with respect to sHCA.
A critical output in quantile-resolved HCAs is the HC value corresponding to a specific, tolerable likelihood level e.g., the 5\,\% quantile.
Given two differing \(\hat{\text{HC}}_{n,x}\) and \(\text{HC}_{n,x}\) corresponding to this likelihood for the true distribution and the approximation.
Then using the approximate value in further calculations, e.g., for grid planning decisions, would lead to a different likelihood of exceeding the operational limits than expected.
Normalization of \(\text{qRE}_{n,x}\) by the standard deviation \(\sigma[\text{HC}_n]\) of the sHCA distribution is crucial, as the sensitivity of the likelihood to HC differences directly depends on the dispersion of the distribution.
A broad HC distribution with a high \(\sigma[\text{HC}_n]\) implies a small change in associated likelihood for the described error, indicating less sensitive likelihood dependency.
Conversely, a narrow distribution with low \(\sigma[\text{HC}_n]\) implies a large change in likelihood for the same error in HC, indicating high sensitivity.
By scaling the absolute quantile HC error by \(\sigma[\text{HC}_n]\), \(\text{qRE}_{n,x}\) transforms the absolute HC error into a measure of likelihood change.
In the context of comparing the true distribution from sHCA to the approximated distribution from pHCA, this under- or overestimation in likelihood is also relevant for using the result e.g., in grid planning.

\section{Results}
\label{sec:results}

In this section, we present the results of applying the HC calculation methods presented in Section II to two exemplary distribution networks.
The loads in the grids are equipped with time series profiles for consumption and generation behavior.
These are selected from a data set that contains real, time-resolved power measurements over one year with a temporal resolution of 15 minutes.
The original dataset is available in \cite{becker_2025}.
Computational effort is evaluated on a MacBook Pro with Apple M3 Pro chip and 24\,GB of RAM.
Important to note is that the sHCA is implemented in Julia with \textit{PowerModels.jl} \cite{coffrin_2018}, while the pHCA is implemented with a built-in OPF adaption from PandaPower in Python.
This introduces a shift in favor of the sHCA, because the performance of Julia is generally better than Python.

\subsection{10-bus Distribution System}
\label{subsec:results-10-bus-distribution-system}

The first network is from the open-source grid modeling library PandaPower \cite{thurner_2018a}, which comprises 4 loads and 10 nodes, 1 transformer and 8 lines.
The power system is named \textit{four\_loads\_with\_branches\_out()}.
The network's line lengths are changed to the values \(l_{4}=0.1\)\,km, \(l_{5}=1.0\)\,km, \(l_{6}=0.05\)\,km, and \(l_{7}=0.3\)\,km.
Lengths and impedances of the lines in the original network do not vary.
With this modification, different behaviors for the optimization and the HC across the different loads can be observed.
We analyze two load sampling approaches:
\begin{enumerate}
    \item \textbf{Normalized Sampling}\\
    For validation purposes, we fit a multivariate normal distribution to the load profiles and use the expected values and covariances as input for the probabilistic optimization approach and draw samples from these normal distributions for the sHCA. 
    \item \textbf{Direct Sampling}\\
    For the comparison of the method including the fit to a multivariate normal distribution, we use the original profile samples taken from the true load distribution as input for the sHCA.
    This result is compared to the results for the pHCA.
\end{enumerate}

\begin{figure}[b]
    \includegraphics[width=\linewidth]{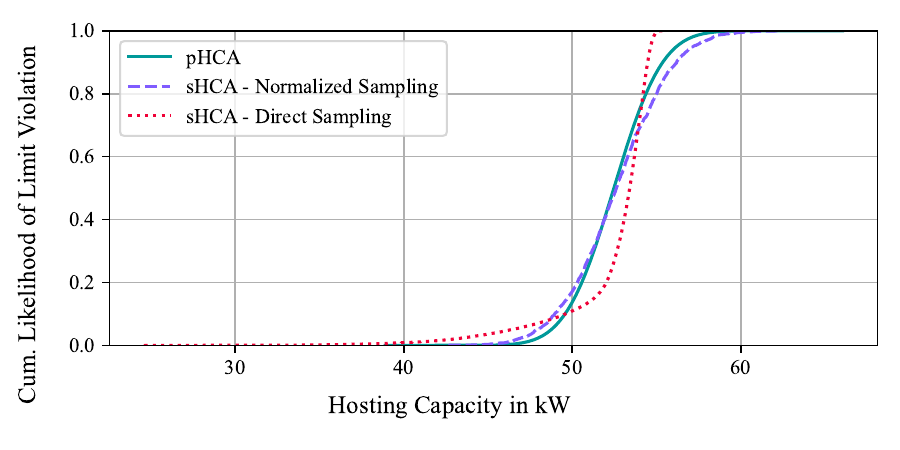}
    \caption{Comparison of the results for the hosting capacity assessment for the 10-bus test system. Shown are the results for both the sHCA considering direct and normalized sampling and the pHCA for the node at which load 1 is connected to.}
    \label{fig:validation-simple-four-bus}
\end{figure}

For the validation with the normalized sampling as input, the results are shown in Fig.~\ref{fig:validation-simple-four-bus}.
In purple, the sHCA approach is shown.
The blue curve shows the results for the probabilistic optimization approach.
For better visibility, only the HC at bus one is shown, as the results for the other nodes are similar.
The interpretation of the result shown in Fig.~\ref{fig:validation-simple-four-bus} reads:
\begin{quote}
    For the analyzed bus, a Hosting Capacity of approx. 50\,kW can be added at 20\,\% probability of exceeding the operational limits.
\end{quote}

\begin{figure}[t]
    \includegraphics[width=\linewidth]{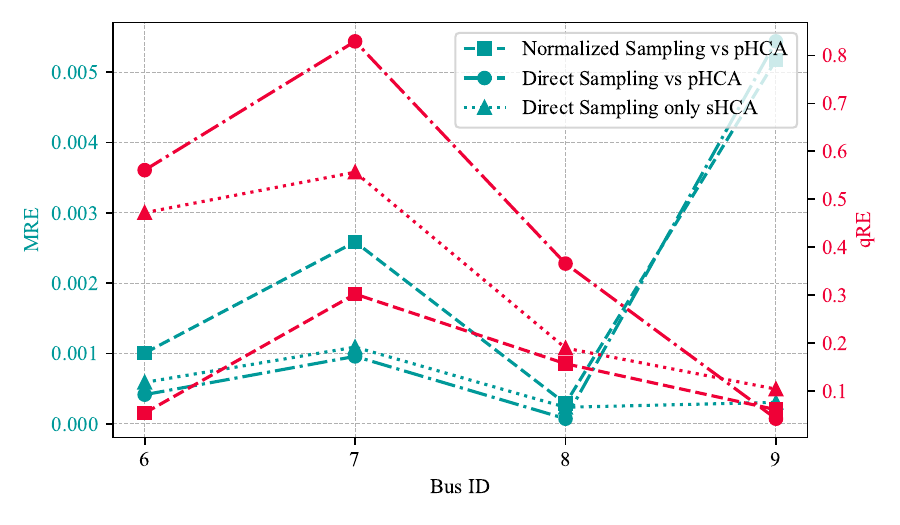}
    \caption{Evaluation metrics of the validation results for the 10-bus test system, when comparing the distributions from the sHCA and the pHCA. Square dashed represents normalized sampling, circle dashdotted represents direct sampling compared with the pHCA. Triangle dotted compares the two scenarios of the sHCA with each other. The plot shows all nodes with connected loads.
    }
    \label{fig:similarity-metrics}
\end{figure}

To compare the pHCA and the sHCA with the direct sampling as input, the results are shown in red in Fig.~\ref{fig:validation-simple-four-bus}.
To account for the true differences between direct and normalized sampling, a comparison of both using sHCA is shown in purple as well.
The metrics evaluating the similarity for normalized as well as direct sampling and the pHCA, are depicted in Fig.~\ref{fig:similarity-metrics}.

\subsection{Demonstration on a Realistic LV Feeder}
\label{subsec:demo-real-grid}

We also analyze the performance of the pHCA when applied to a real LV from a German DSO.
The network consists of 105 nodes, 38 loads, 102 lines, and 1 transformer.
For this network, we also assess the hosting capacity by applying the load sampling methods introduced above.

\section{Discussion}
\label{sec:discussion}

First, the results show consistency between the studied approaches, with the pHCA closely matching the sHCA for all nodes and quantiles when looking at the obtained CDFs.
Slight deviations can be observed e.g., a slight offset or shift of the distributions.
Another observation are the differences in the tails of the obtained HC distributions, as the plot shows different distribution widths for the pHCA and the sHCA.
Possible explanations for these differences can be the multivariate normal distribution fitting, which is not able to capture the tails of the original load distribution.
Additionally, the linearization of the power flow model can lead to errors in the state distribution calculation and thus in the constraint reformulation.
Further, covariances between the load hypothesis and the added DER capacity are not considered in the formulation of the OPF in Eq.~(\ref{eq:pgm-optimization}), which can lead to errors in the results as well.
These mentioned sources of errors influence especially the tails of the HC distribution.

\begin{figure}[t]
\centering
        \includegraphics[width=\linewidth]{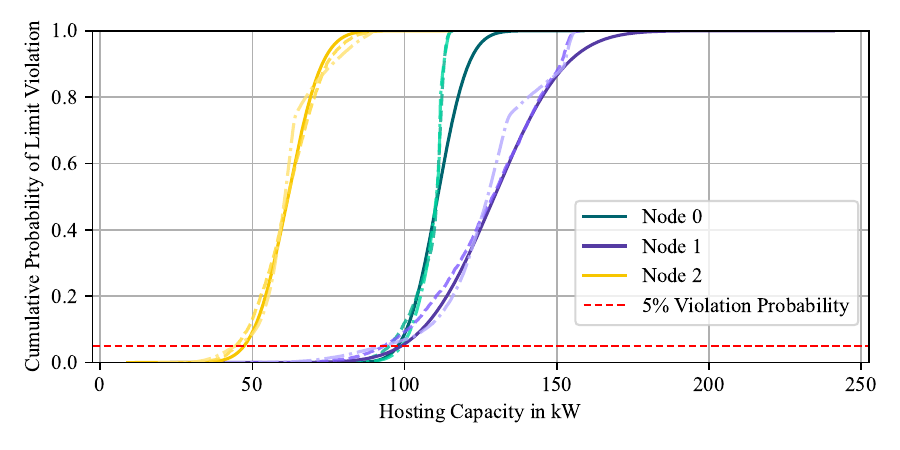}
\label{fig:results-pgm-realistic-lv-feeder}
\caption{
      Results for the hosting capacity assessment for the realistic LV feeder.
      Selected nodes are compared with their CDF functions in the respected color family.
      Dark represents the probabilistic optimization approach, light equals to direct sampling, and the middle normalized sampling.
      The results show the 5\,\% quantile of the HC distribution for all nodes as CDF. }
\end{figure}

When looking at the similarity indices, the \(\text{MRE}\) for all three comparisons is below or around 0.3\,\% for the simple 10-bus grid, which indicates a good agreement between the pHCA and the sHCA in terms of the expected values of the HC distributions.
Only bus 9 has an elevated error of around 0.5\,\%. 
The \(\text{qRE}\) is below 0.6, except the raise to around 0.8 for the direct sampled comparison at node 7.
Comparing node 7 and 9, the combination of both indices indicate a tilt of the distribution for bus 7, and a tilt combined with an offset for bus 9.
Overall the obtained quantiles of the HC distributions are consistent.
For the comparison considering direct sampling against normalized sampling on the sHCA, the qRE is approx. 0.1 to 0.5.
As this quantifies the true differences coming from the normal distribution fitting, it is interpreted as systematic offset.
This is expected as the original profiles are usually not well representable by normal distributions.
It has to be considered for the evaluation of the comparative metric for the pHCA and direct sampling.
When applying the individual offsets for the direct sampling comparison of the pHCA, the absolute value of the qRE comes down to approx. the same level as for the normalized sampling comparison.
This supports the assumption of the normal distribution fitting being a significant source of error in the HC results, quantified by the triangled dotted data in Fig.~\ref{fig:similarity-metrics}.

Considering the results from the application on the realistic LV feeder, the results show a similar behavior in calculation and appearance as for the 10-bus system.
The MRE is not exceeding 3.5\,\%, while the qRE is below 0.8 for almost all assessed nodes. 
Here the proposed HC calculation method shows its strengths, as results keep a stable error while the computational effort scales very beneficial for e.g., large number of samples or fast direct application in operational applications.

Regarding the run time of the two approaches, the sHCA takes about 11\,min for the calculation of the HC for the complete 10-bus network within a load hypothesis of around 35k time steps, representing a year in quarterly-hour granularity.
The pHCA takes about 2\,s for the identical task.
The run time for the realistic LV feeder within the same setup is about 50\,s for solving two instances of Eq.~(\ref{eq:pgm-optimization}), and about an estimated 38\,h for generating results with the sHCA considering a sample set of 35k steps.

The faster and more efficient pHCA is promising for large networks, long time intervals, or other applications with a demand on fast calculations.
Especially if the load hypothesis can be well represented by normal distributions, the method can provide results within 3-4\,\% error on realistic LV grids, with a significant reduction in computational effort.
This is the case when considering the integration of other uncertainties in the method, such as grid parameters, measurement errors, or forecasts.

\section{Summary and Outlook}
\label{sec:summary}

This paper introduces an approach for quantile-resolved hosting capacity assessment, which is an efficient alternative to the sHCA approach.
Our simulation experiments on two test networks show consistency between the two approaches, with the pHCA closely matching the sHCA for all nodes and quantiles.
Computational effort is significantly reduced with the probabilistic optimization approach, which makes it a promising method for large networks, large number of samples, or other applications with a demand for repeated fast calculations.

Future work will address the extension and application of the pHCA in pro-active operational planning for the reduction of grid congestions, and its applicability and integration in curative LV grid or flexibility management solutions.
Another aspect of future work is the extension of the method to other load distributions, e.g., Gaussian Mixture Models, which can provide a better representation of the original load behavior and thus further increase accuracy in the results of the compared methods from the multivariate normal distribution fitting.
Further developments will focus on the adequate assessment of flexibility representation in HCA methods, as these can provide significant benefits for the integration and management of DERs.
These methods can also support DSOs in planning decisions like grid connection assessment.

\printbibliography

\end{document}